\documentstyle[aps,prl,epsf]{revtex}
\begin{document}
\twocolumn[\hsize\textwidth\columnwidth\hsize\csname@twocolumnfalse\endcsname
\author{A.E.Koshelev}
\address{Material Science Division, Argonne National Laboratory,Argonne, IL 60439,
and\\ 
Institute of Solid State Physics, Chernogolovka, Moscow District, 142432, Russia}
\title{Plasma resonance and remaining Josephson coupling in the 
``decoupled vortex liquid phase'' in layered superconductors.}
\date{\today }
\maketitle
\begin{abstract}
We relate the frequency of the Josephson plasma resonance in layered 
superconductors with the frequency dependent superconducting response.  
We demonstrate that the sharp resonance can persist even when the 
global superconducting coherence is broken provided the resonance 
frequency is larger than the frequency of interlayer phase slips.  In 
this situation the plasma frequency is determined by the average 
Josephson energy, which can be calculated using the high temperature 
expansion.  We also find the temperature dependence of the average 
Josephson energy from the Monte Carlo simulations and determine the 
applicability region of the high temperature expansion.  
\end{abstract}
\twocolumn 
\vskip.2pc] 
\narrowtext
Interlayer Josephson coupling between superconducting layers has 
crucial influence on thermal and pinning properties of layered 
superconductors in the mixed state.  A new, powerful tool to probe the 
Josephson coupling in a wide range of temperatures and fields has been 
introduced recently.  A sharp microwave absorption resonance has been 
discovered in the vortex state of $ Bi_2Sr_2CaCu_2O_8$ 
\cite{tsui1,tsui2,matsuda}, which was shown to be caused by electric 
field across the layers
\cite{matsuda} and was attributed to the earlier predicted plasmon 
resonance \cite{Artemenko,tachiki,bmt,pokrovsk}.  It was found that the 
resonance can be tuned by temperature and magnetic field.  The 
resonance field corresponding to the fixed microwave frequency of 30-90 
GHz increases with temperature at low temperatures having a maximum at 
the irreversibility temperature and afterwards smoothly decreases with 
temperature.  The resonance frequency at a given temperature and field 
can serve as a measure of the effective Josephson coupling.  
Unexpectedly, the plasmon resonance has been observed up to high temperatures 
$\sim$75 K \cite {matsuda} where the Josephson coupling should 
be strongly suppressed by thermal fluctuation of pancake vortices.  It 
was theoretically predicted that pancake fluctuations should actually 
destroy the global superconducting coherence above the decoupling line 
\cite{GlazKosh,Daemen}.  Matsuda {\it et al} \cite{matsuda} suggested 
that the decoupling line, if it exists, should lie above the 
temperature-field line at which the sharp plasmon resonance at a given 
microwave frequency is observed.  However, their line of resonance, $ 
B_0(T)$, clearly lies above the decoupling line observed by $I-V$ 
measurements \cite{Cho} and falls into the region where linear 
resistivity in the c-direction was observed 
\cite{Cho,Briceno,Latyshev,Busch}.  In this Letter we resolve this 
controversy.  We demonstrate that the sharp plasmon resonance can 
actually persist even above the decoupling transition provided the 
resonance frequency is higher than the typical frequency of the 
interplane phase slips.  In this situation the plasma frequency is 
determined by the average Josephson energy $E_J^{eff}=E_J\left
\langle \cos \phi _{nn+1}\right\rangle$ (or, equivalently, the average 
Josephson current) as was suggested in Refs.~\cite{tsui2,matsuda}.  
Here $\phi _{nn+1}=\phi _{n+1}-\phi _n-\frac{2\pi s}{\Phi _0 }A_z$ is 
the gauge invariant phase difference between layers and $E_J$ is the 
bare Josephson energy.  The decoupling transition manifests itself by 
a disappearance of the {\it linear superconducting response} along the 
{\it c}-axis at zero frequency while the average Josephson energy 
should not change much at this point and it actually stays finite at 
any temperature.

The superconducting response, $Q_\omega $, determines the response of 
the electric current to the external alternating vector-potential, 
$A_\omega $ \cite{textbook}
\begin{equation}
j_\omega =-Q_\omega A_\omega.   \label{ScResp}
\end{equation}
In superconducting state $Q_{\omega \rightarrow 0}>0$.  For the 
layered superconductor with the Josephson energy ${\cal E}_J=-E_J\int 
d{\bf r}\sum_n\cos \phi _{nn+1}$, one can obtain a general thermodynamic 
expression for the z-component of $Q_\omega $ using the 
fluctuation-dissipation theorem (FDT)(see, e.g., 
Ref.~\onlinecite{StatPhys}).  The external vector-potential, $ A_\omega $, 
creates the energy perturbation, ${\cal E}_A=-\frac{j_J}c\int d {\bf 
r}\sum_n\sin \phi _{nn+1}A_\omega $, with $j_J=\frac{2\pi c}{\Phi 
_0}E_J$ being the Josephson current.  The average current along 
the z-direction $ j_z=j_J\left\langle \sin \left(
\phi _{nn+1}-\frac{2\pi sA_\omega }{\Phi _0} \right) \right\rangle 
_\omega $ in linear with respect to $A_\omega $ approximation is given 
by
\begin{equation}
j_z=j_J\left( -\left\langle \cos \phi _{nn+1}\right\rangle _0\frac{2\pi
sA_\omega }{\Phi _0}+\left\langle \sin \left( \phi _{nn+1}\right)
\right\rangle _\omega \right)   \label{jz}
\end{equation}
Here $\left\langle ...\right\rangle _0$ notates the thermodynamic 
average with undisturbed energy and $\left\langle ...\right\rangle 
_\omega $ notates averaging which takes into account the perturbation 
${\cal E}_A$.  The second term describes the average Josephson 
current, which emerges due to the phase perturbations caused by 
the external field.  Using the FDT relation \cite{StatPhys} this term 
can be connected with the correlation function of the Josephson 
currents and we obtain the following expressions for the real 
($Q_\omega ^{\prime }$ ) and imaginary ($Q_\omega ^{\prime \prime }$) 
parts of the superconducting response along the $z$-direction
\begin{eqnarray}
Q_\omega ^{\prime } &=&sj_J\left[ \frac{2\pi }{\Phi _0}\left\langle \cos
\phi _{nn+1}\right\rangle +\frac{j_J}{cT}\int d{\bf r}\int\limits_0^\infty
dt\sum_n\right.   \label{ScRespRe} \\
&&\left. \cos (\omega t)\frac d{dt}\left\langle \sin \phi _{01}({\bf 0}
,0)\sin \phi _{nn+1}({\bf r},t)\right\rangle \right]   \nonumber \\
Q_\omega ^{\prime \prime } &=&-\frac{\omega sj_J^2}{cT}\int d{\bf r}
\int\limits_0^\infty dt\sum_n  \nonumber \\
&&\cos \left( \omega t\right) \left\langle \sin \phi _{01}({\bf 0},0)\sin
\phi _{nn+1}({\bf r},t)\right\rangle   \label{ScRespIm}
\end{eqnarray}
Substituting the material relation (\ref{ScResp}) into the Maxwell 
equations one can see that the plasma resonance exists at the 
frequency $\omega _{pl}$ connected with $Q_\omega^{\prime }$ by relation 
\cite{pokrovsk}
\begin{equation}
\omega _{pl}^2=4\pi cQ_\omega ^{\prime }/\epsilon   \label{PlasmFreq}
\end{equation}
provided $Q_\omega ^{\prime }\gg Q_\omega ^{\prime \prime }$. 
Here $\epsilon$ is the dielectric constant.  Due to the
frequency dependence of $Q_\omega ^{\prime }$ the last expression is
actually an equation for $\omega _{pl}$ rather than its definition. 
As one can see from Eq.(\ref{ScRespRe}) $Q_\omega ^{\prime }$ is given 
by the sum of two terms, the first is determined by the average 
Josephson current, $ j_J\left\langle \cos \phi _{nn+1}\right\rangle $, 
and the second is determined by the correlation function of the 
Josephson currents.  In the decoupled liquid phase $Q_{\omega 
=0}^{\prime }=0$, which means that at $\omega =0$ the second term in 
square brackets of Eq.(\ref{ScRespRe}) exactly compensates the first 
one.  The frequency dependence is totally determined by the second 
correlation term.  This term is strongly suppressed when the frequency 
exceeds the typical ``phase slip'' frequency $\omega _{ps}$, which is 
determined by the typical decay time of the correlation function $
\left\langle \sin \phi _{nn+1}({\bf 0},0)\sin \phi _{nn+1}({\bf 0}
,t)\right\rangle $. In the high frequency regime, $\omega \gg \omega 
_{ps}$,
we have $Q_\omega ^{\prime }\approx \frac{2\pi }{\Phi _0}sj_J\left\langle
\cos \phi _{nn+1}\right\rangle $, i.e., the plasma resonance probes the 
average Josephson current.  In the mixed state $\left\langle \cos \phi 
_{nn+1}\right\rangle $ is suppressed by misalignment of pancake vortices 
induced by pinning and thermal fluctuations.  In Ref.\onlinecite{BPM} this 
quantity has been connected with the density correlation functions assuming 
Gaussian distribution for $\phi _{nn+1}$.

In the decoupled liquid phase $\left\langle \cos \phi _{nn+1}\right\rangle 
$ can be calculated using the high temperature expansion with respect to the 
Josephson energy which gives
\begin{equation}
\left\langle \cos \phi _{nn+1}\right\rangle =\frac{E_J}{2T}\int d{\bf r}
\left| S_1({\bf r})\right| ^2,  \label{HighTExp}
\end{equation}
where 
\begin{equation}
S_1({\bf r})=\left\langle \exp \left[ i\left( \phi ({\bf r})-\phi ({\bf 0})-
\frac{2\pi }{\Phi _0}\int_{{\bf 0}}^{{\bf r}}d{\bf lA}\right) \right]
\right\rangle   \label{PhaseCorr}
\end{equation}
is the phase correlation function for a single 2D layer.  In the vortex 
liquid state the decay length of $S_1({\bf r})$ is given by the average 
intervortex spacing and Eq.  (\ref{HighTExp}) can be rewritten as
\begin{equation}
\left\langle \cos \phi _{nn+1}\right\rangle =\frac{E_J\Phi _0}{2TB}f,
\label{HighTExp1}
\end{equation}
where $f=\frac B{\Phi _0}\int d{\bf r}\left\langle \left| S_1({\bf 
r})\right| ^2\right\rangle $ is the dimensionless function of order unity 
with weak temperature dependence.  We find from simulations 
that the the high-T expansion is quantitatively accurate if $\left\langle \cos \phi 
_{nn+1}\right\rangle\lesssim 0.5$.  This corresponds to the field and 
temperature range $TB > \Phi _0E_J$. Combining Eqs.~(\ref{HighTExp1})
and (\ref{PlasmFreq}) we obtain
\begin{equation}
\omega _{pl}^2=\frac{2\pi fsj_J^2\Phi _0}{\epsilon TB}.  \label{PlHighT}
\end{equation}
Taking into account the weak temperature dependence of $f$, this expression 
gives an explanation for the experimentally observed dependence $\omega 
_{pl}^2\propto 1/TB$ \cite{matsuda}.  Note that such scaling indicates 
almost decoupled layers rather than a persistence of coupling at high 
temperatures.

The magnitude and shape of the function $f(T)$ depend upon the physical 
realization of the 2D liquid state in the layers.  In pin free 
superconductors this function is universal and depends only on the 
dimensionless parameter $T(4\pi \lambda _{ab})^2/s\Phi _0^2$.  Below we 
calculate this universal function using Monte Carlo simulations.  In real 
samples random pinning increases disorder in vortex arrangements even in 
the liquid state, which leads to extra suppression of $f(T)$ and further 
smoothes its temperature dependence.  On the other hand, the pinning effect 
becomes progressively weaker at higher temperatures.

Substituting (\ref{HighTExp}) into (\ref{ScRespRe}) we obtain the high-T
expansion for $Q_\omega^\prime $
\begin{equation}
Q_\omega ^{\prime }=\frac{\omega sj_J^2}{2cT}\int d{\bf r}
\int\limits_0^\infty dt\sin \left( \omega t\right)  
S({\bf r},t),  \label{ScReHighT}
\end{equation}
with $ S({\bf r},t)=\left\langle \exp \left[ i\left( \phi_{nn+1} ({\bf 
r},t) -\phi_{nn+1} ({\bf 0},0)\right) \right]\right\rangle $ being the 
dynamic phase correlation function.  One can immediately see that, indeed, 
in this regime $Q_{\omega =0}^{\prime }=0$, indicating decoupled layers.

The result (\ref{HighTExp1}) can be generalized to the case when the 
magnetic field is tilted with respect to the $c$-axis.  In presence of the 
$y$-component of the magnetic field, $B_{y}$, and in the lowest order with 
respect to $E_J$, the phase difference acquires an extra contribution $\phi 
_{nn+1}\rightarrow \phi _{nn+1}+\frac{2\pi s}{\Phi _{0}}B_{y}x$, which 
gives an extra factor $\exp (i\frac{2\pi s}{\Phi _{0}}B_{y}x)$ in the RHS 
of Eq.~(\ref{HighTExp}).  Approximating $S_{1}({\bf r)}$ by a Gaussian 
function $S_{1}({\bf r)}\approx
\exp (-\frac{\pi Br^{2}}{2f\Phi _{0}})$ we obtain
\begin{equation}
\left\langle \cos \phi _{nn+1}\right\rangle =\frac{fE_{J}}{2T}\exp (-f\frac{
\pi s^{2}B_{y}^{2}}{B_{z}\Phi _{0}})  \label{cosHyAppr}
\end{equation}
The Josephson coupling in tilted fields in presence of coexistent 
Abrikosov and Josephson lattices was studied in 
Ref.~\onlinecite{BulHpar}.  As follows from Eq.~(\ref{cosHyAppr}) the 
typical $ab$-component of the field is given by $\sqrt{{B_{z}\Phi 
_{0}}/{\pi s^{2}}}$ and coincides with the one found in 
Ref.~\onlinecite{BulHpar} at high fields 
$B_{y}>\Phi _{0}/\gamma s^{2}$.  For $B_z=100$ G this typical field is 
$\sim$ 1.7 T which corresponds to a tilt angle $\sim$ $0.3^\circ$.  
In contrast to the lattice state, the average Josephson current does 
not experience oscillations caused by commensurability of Abrikosov and 
Josephson lattices but, instead, decreases monotonically with $B_y$.

One can expect that in the liquid state the phase slips are strongly 
assisted by thermal motion of pancake vortices.  In real samples this 
motion is strongly hindered by pinning as follows from the 
thermally activated behavior of the in-plane component of resistivity 
$\rho _{ab}$\cite{rho-ab,Busch} in a wide temperature range.  An important 
experimental fact is that the temperature dependencies of $\rho _{ab}$ 
and $\rho _c $ are characterized by the same value of the activation 
energy \cite {Latyshev,rho_c-model,Safar}.  This indicates that the 
interlayer phase slips are, indeed, induced by motion of pancake 
vortices.  These phase slips are very well described by a simple model 
which assumes that the dissipation processes are caused by a small 
concentration $n$ of mobile pancakes with a thermally activated 
diffusion constant $D$ \cite{rho_c-model}.  Within this model the 
dynamic phase correlation function decays in time and space as
\begin{equation}
S(r,t)=\exp \left( -\frac{\pi Br^2}{2\Phi _0}\ln \frac{r_{{\rm cut}}}r-\pi
nDt\ln \frac{r_{{\rm cut}}^2}{Dt}\right) ,  \label{Srt-model}
\end{equation}
where $r_{{\rm cut}}$ is the in-plane phase coherence length.
Substituting the last equation into Eq.~(\ref{ScReHighT}) we obtain 
\[
Q_\omega ^{\prime }\approx \frac{sj_J^2\Phi _0}{cTB}\frac{\omega ^2}{\omega
^2+\omega _{ps}^2},
\]
where the typical phase slip frequency, $\omega _{ps}$, is given by $\omega 
_{ps}=2\pi nD\ln \left( nr_{{\rm cut}}^2\right) $.  Because the same mobile 
pancakes determine the dissipation along the {\it ab}-plane, this frequency 
can be connected with $\rho _{ab}$ \cite{rho_c-model}
\begin{equation}
\omega _{ps}=2\pi \ln \left( nr_{{\rm cut}}^2\right) \left( \frac c{\Phi _0}
\right) ^2\frac Ts\rho _{ab},  \label{omega-rho}
\end{equation}
or $\omega _{ps}\approx 2\div 4\cdot 10^8$[1/s]$\cdot T$[K]$\cdot \rho _{ab}$
[$\mu \Omega \cdot $cm]. Taking $T=45$ K and $\rho _{ab}=10^{-3}\mu \Omega
\cdot $cm corresponding to the resonance field $B\approx 0.1T$ \cite{matsuda}
we obtain $\omega _{ps}\approx 1\div 2\cdot 10^7$[1/s], which is indeed much
smaller than the microwave frequency $\omega =2\pi \cdot 4.5\cdot 10^{10}$
[1/s] used in Ref.\cite{matsuda}. As follows from 
Eqs.~(\ref{ScRespIm},\ref{Srt-model}) $Q_\omega ^{\prime \prime}\approx 
(\omega_{ps}/\omega)Q_\omega ^{\prime }$ at $\omega\gg \omega _{ps}$, i.e. 
the plasma resonance is not suppressed by the dissipation in this regime.
Eq.~(\ref{omega-rho}) holds only in the ``phase slip'' regime, i.e., until 
the quasiparticle contribution to the c-axis transport can be neglected.

The analytical calculation of the correlation function 
(\ref{PhaseCorr}) in the liquid state is not possible even for a pin-free 
superconductor.  To obtain quantitative information about the Josephson 
coupling in the liquid state and to check the high-T expansion we perform 
Monte Carlo simulations of the frustrated three dimensional XY model which 
was used to investigate phase transitions in the vortex state
\cite{LiTeitel,Hetzel}. The energy of the model is given by
\begin{equation}
E=\sum_{{\bf n},\alpha }V_\alpha \left( \phi ({\bf n+\delta }_\alpha )-\phi (
{\bf n})-a_\alpha ({\bf n})\right)   \label{EnergXY}
\end{equation}
Here $\phi ({\bf n})$ are the phases of the order parameter located at the 
sites of a mesh $1\leq n_x,n_y\leq N_{\perp }$,$1\leq n_z\leq N_z$, with 
periodic boundary conditions in all directions.  The vector potential 
$a_\alpha ({\bf n})$ ($\alpha =x,y,z$ ) is taken in the Landau gauge 
$a_y(n_x)=2\pi bn_x$, where the dimensionless magnetic field $b$ determines 
the fraction of the grid sites occupied by vortices.  We chose $ N_{\perp 
}=72$, $N_z=4,10$, and $b=1/36$ which gives $144$ vortex lines.  The 
in-plane phase-phase interaction function, $V_x(\theta )=V_y(\theta )$, 
determines the in-plane phase stiffness.  The choice of the low angle 
asymptotics, $V_{x,y}(\theta )\approx -1+\theta ^2/2$, fixes the energy and 
temperature scale as $s\Phi _0^2/\left( \pi \left( 4\pi \lambda
\right) ^2\right) $.  An unrealistic feature of the 
model is that vortices are subject to the periodic pinning potential 
created by the numerical grid.  This effect can be reduced significantly by 
optimizing the shape of $V_x(\theta )$.  We chose $V_x(\theta 
)=-3r/4-(1-r)\cos (\theta )-(r/4)\cos (2\theta )$, and adjusted the 
coefficient $r$ to minimize the energy barrier for the vortex jump to the 
neighbor site, which gives $r=0.376$.  The coupling between planes is given 
by the standard Josephson interaction, $V_z(\theta )=-\frac 1\Gamma \cos 
\left( \theta \right) $, with the amplitude determined by the anisotropy 
parameter $\Gamma $.  The coupling between the two dimensional lattices in 
the layers is determined by a scaled magnetic field $h=B/B_{cr}$, where 
$B_{cr}=\Phi_0/(\gamma s)^2$ is the crossover field \cite{GlazKosh} and 
$\gamma$ is the anisotropy of the London penetration depth (in 
dimensionless units $h=\Gamma b$).  The parameter $\Gamma$ should not be 
considered as the anisotropy of a real layered superconductor but rather as 
a parameter which allows us to vary the scaled magnetic field $h$.  We 
apply the standard Metropolis algorithm to the Hamiltonian (\ref{EnergXY}).  
The evolution of the phase diagram with changing anisotropy will be 
reported elsewhere.  In this Letter we will focus on the calculation of the 
quantity $\left\langle \cos \phi _{nn+1}\right\rangle$ and verification of 
the high-T expansion for this parameter.

\begin{figure}
\epsfxsize=3.2in 
\epsffile{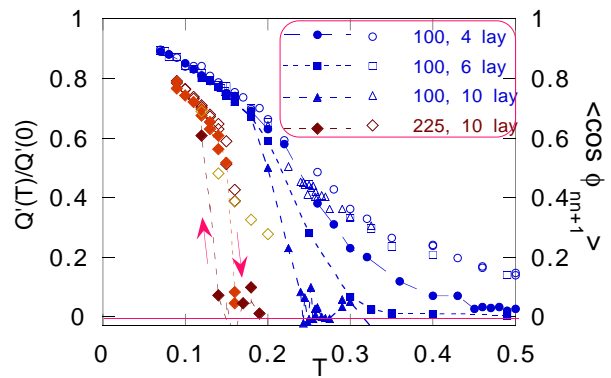}
\caption{Temperature dependencies of the superconducting response at 
$\omega=0$ (filled symbols) and parameter $\left\langle \cos \phi 
_{nn+1}\right\rangle$ (open symbols) for two values of the anisotropy parameter 
$\Gamma=100$ and $225$}
\label{Fig-Q-cos}
\end{figure}
Fig.~\ref{Fig-Q-cos} shows the numerically obtained temperature 
dependencies of the low frequency superconducting response 
$Q^\prime=Q^\prime_{\omega=0}$ and the parameter $\left\langle \cos \phi 
_{nn+1}\right\rangle$ for two values of the anisotropy ($\Gamma=100$ and 
$225$).  As one can see from this plot the temperature dependencies of 
these parameters are almost identical above the level $\left\langle \cos 
\phi _{nn+1}\right\rangle=0.6$.  Below this level $Q^\prime(T)$ starts to 
drop fast, indicating the decoupling transition, and $\left\langle \cos 
\phi _{nn+1}\right\rangle$ still decreases smoothly.  For high anisotropies 
$\left\langle \cos \phi _{nn+1}\right\rangle$ has a small drop at the 
decoupling transition.  One can see also that for $\Gamma=100$, 
$Q^\prime(T)$ shows a strong size effect for $n_z<10$, which indicates that 
this parameter is not simply determined by the interaction of two neighbor 
layers but rather characterizes the superconducting coherence along the 
whole sample thickness.

\begin{figure}
\epsfxsize=3.2in 
\epsffile{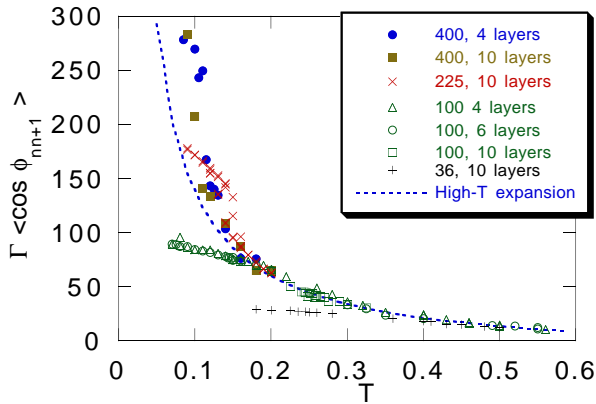}
\caption{Temperature dependencies of the parameter $\Gamma \left\langle \cos \phi 
_{nn+1}\right\rangle$ for different $\Gamma$. High-T expansion curve for 
this parameter given by Eq.~(\protect{\ref{HighTDless}}) is also plotted. }
\label{Fig-Gam-cos}
\end{figure}

To verify the high-T expansion we calculate the phase correlation 
function (\ref{PhaseCorr}) at different temperatures and integrate it to 
determine the universal function $f(\tilde{T})$ in Eq.~(\ref{HighTExp1}).  
The last equation takes the following form in dimensionless units
\begin{equation}
\label{HighTDless}
\left\langle \cos \phi 
_{nn+1}\right\rangle=f(\tilde{T})/(2\Gamma\tilde{T}b).
\end{equation}
We found that the temperature dependence of $f(\tilde{T})$ is, indeed, 
rather weak and can be approximated by the formula 
$f(\tilde{T})=0.862-0.976\;\tilde{T}$ for $0.1<\tilde{T}<0.6$.  For 
parameters corresponding to BiSSCO, this interval covers the 
temperature range $24$ K $<T<66$ K.  In Fig.~\ref{Fig-Gam-cos} we plot 
the temperature dependencies of the directly calculated $\Gamma
\left\langle \cos \phi _{nn+1}\right\rangle$ at different values of 
$\Gamma $ together with it's high-T expression (\ref{HighTDless}).  As 
one can see from this plot, all curves merge with the same universal curve 
at high enough temperatures which is very well approximated by the 
high-T expansion.  Systems with lower anisotropy merge with this 
curve at higher temperatures.

We can use these results for quantitative analysis of the plasma 
resonance data at high temperatures.  Using the relation 
$j_J=2c\Phi_0/\left[ s (4\gamma\pi \lambda_{ab})^2\right]$ one can 
extract the anisotropy ratio $\gamma$ from Eq.(\ref{PlHighT}).  Taking 
the resonance field $B_0=0.065$ T at $T=53.9$ K and $\omega/2\pi=45$ 
GHz from Ref.~\onlinecite{matsuda} and using the material parameters 
$\epsilon=20$ and $\lambda_{ab}(T)=1800\AA/\sqrt{1-(T/T_c)^2}$ we 
calculate from the simulation results 
$\left\langle \cos \phi _{nn+1}\right\rangle\approx 0.11$ and 
$\gamma\approx 330$. 

This work was initiated by discussions with L.~Bulaevskii.  Very useful 
discussions with M.~Tachiki, Y.~Matsuda, and M.~Gaifullin are also 
appreciated.  The work was supported by the National Science 
Foundation Office of the Science and Technology Center under contract 
No.  DMR-91-20000.  and by the US Department of Energy, BES-Materials 
Sciences, under contract No.  W-31- 109-ENG-38.  The author gratefully 
acknowledges use of the Argonne High-Performance Computing Research 
Facility.  The HPCRF is funded principally by the U.S.  Department of 
Energy Office of Scientific Computing.

\end{document}